\documentclass[a4paper]{article}
\usepackage{graphicx}
\newcommand{\be}{\begin{equation}}
\newcommand{\ee}{\end{equation}}
\newcommand{\bd}{\begin{displaymath}}
\newcommand{\ed}{\end{displaymath}}
\newcommand{\bea}{\begin{eqnarray}}
\newcommand{\eea}{\end{eqnarray}}
\newcommand{\bi}{\begin{description}}
\newcommand{\ei}{\end{description}}
\newcommand{\bq}{\begin{quote}}
\newcommand{\eq}{\end{quote}}

\def\fo{\footnote}

\begin{document}
\bibliographystyle{unsrt}
\setcounter{secnumdepth}{0}
\twocolumn[
\author{Alexander~Unzicker\\
        Pestalozzi-Gymnasium  M\"unchen\\[0.6ex]}

\title{The VSL Discussion: What Does Variable Speed of Light Mean and Should we be Allowed to Think About~?}
\maketitle

\begin{abstract}
In the past years, variable speed of light (VSL) theories have
been of growing interest but also a subject of controversial
discussion. They have been accused both for tautologies and for
violating special relativity, and concerns have been expressed
about the validity of such approaches in general (e.g. Ellis,
astro-ph/0703751). Without trying  completeness on the issue, the
example of Einstein's VSL attempts (1911) and Dicke's
`electromagnetic' theory (1957) are urges to give some comments on
the above criticism.
\end{abstract}
\vspace{1.0cm}]

\subsection{Introduction}

\paragraph{Exotic theories.}
Of course, `variability' can encompass a lot of aspects. One may  introduce dispersion,
considering a dependency on $\lambda$, or on $v$, violating
Lorentz-invariance. Most of these proposals do not have sufficient experimental support
at the moment, though many of them are interesting and seem as good as inflation
for resolving the flatness and horizon problems in cosmology; this 
however is not the focus of interest here, since
comments on \cite{Ell:07} with respect to modern
VSL theories \cite{Mag:03}
have already be given \cite{Mag:07}. There, appropriate reference and a clear discussion of
older attempts are however missing. These so-called \cite{Mag:03}

\paragraph{Conservative theories} suffered an even harsher
`Not even wrong'- criticism of being tautological. 
The argument is the following:
`One assumes a good clock can be constructed, and then uses the timing of reflected
electromagnetic radiation to determine the distance. But then the (physical)
speed of light of necessity has to be unity, precisely because all electromagnetic
radiation travels at the speed of light, and distances are being determined by use
of such radiation.' (\cite{Ell:07}, sec.~2). One may wonder what fact should be proven by that
statement.
All that follows indeed from the definition of SI units, but in my humble opinion something
can either be  measured or defined, not both. Thus $c=1$ is not a physical necessity
but at best a mathematical convention; one may further ask if it is a possible, reasonable
or even the only practical one. At the very end, this is not a scientific question; to illuminate
the practical value of  $c=1$, we investigate the following toy theory:

\subsection{Meteorology at constant temperature}
Fortunately, glass and most fluids have different thermal expansion coefficients (TEC),
and for that reason we easily construct thermometers based on the expansion
of fluids with respect to the containing tube.
But imagine all substances had the same TEC, things wouldn't be that easy!
All thermometers made in that fashion would show the same temperature.
Well, one still could measure temperature by means of the mean quadratic velocities
of particles in a gas. Determining the velocities with clocks and rods and deriving
the temperature would still be  possible. But what if the same velocity is
used for the {\em definition\/} of time and length scales~? A gas thermometer
in a cold location would then just mimic a slower running time and/or a contraction of length scales.
One realizes that in such a world it is not easy to detect temperature differences,
but {\em there is\/} an effect: the velocity of sound waves, depending on temperature,
would be a function of place and time, and hence, differences in temperature would
cause a deflection and focussing of sound waves.

Mind now the following mathematical insight: there cannot be any
doubt that the numerical value of temperature depends on arbitrary
chosen units, and since it is a dimensionful quantity, it can be
set to unity in every point (see argument \cite{Ell:07}, sec.~2).
Therefore, mathematicians should feel free in formulating
meteorology (or, appreciating generalizations, thermodynamics)
with $T=1$, but the demand that any weather forecast should be
expressed in this manner will be of limited usefulness. People who
do not shy elementary material should  have a look
at the textbook example in Feynman's lectures II, chap.~42
\cite{FeyII}\fo{Though being a toy theory, there are very
interesting comments regarding the topic given by Landau
\cite{Lan5}, par.~8.}.

\paragraph{Differential geometry.}
I shall like to draw attention to the fact that such a convention ($\mbox{T=1, c=1}$)
leads to a curved space, which  equivalently can be described by a metric.
 However it is quite a difference if one can choose
an -arbitrary-  unit  {\em globally\/} or if you have to do this {\em locally\/} in every point.
In the later case such a choice $T=1$ can turn out to be complicated. The  `proof' instead
that `physically' $c$ is always a constant is like the
proof of a differential geometer that physically
no mountains exist, since in every point of a differentiable manifold one can attach
a flat tangent space (the necessity to change direction when walking uphill is nothing
physical, just a `connection'.). What a nice revival
of the earth as a plane! We proceed a little further in history and listen to those who first
considered a variable speed of light:

\subsection{Einstein and a VSL.}
The first who realized that a variable speed of light may cause
astronomical light bending was Einstein in 1911 \cite{Einst:11}:
\bq `From the proposition which has just been proved, that the
velocity of light in the gravitational field is a function of the
place, we may easily infer, by means of Huygens's  principle, that
light-rays propagated across a gravitational field undergo
deflexion'. \eq As a consequence of a variable speed of light, he
considered variable time scales only and postulated \be
\frac{dc}{c} = \frac{df}{f}, \ee which, as it is well-known, led
to he (wrong) half value for the classical light deflection.\fo{It
should be noted that though $c$ being a scalar field here, this
theory is not a `scalar' theory {\em coupled to matter\/} to which
Einstein later expressed caveats. See also \cite{Giu:06} for
clarifying that point.}.
 It was then Dicke\fo{It is not quite  clear why Dicke gave up this
interesting approach and followed up
the quite different Brans-Dicke theory. } with his `electromagnetic' theory
of gravitation \cite{Dic:57} who discovered
that the classical tests could be described by
\be
\frac{dc}{c} = \frac{d \lambda}{\lambda} + \frac{df}{f},
\ee
considering variable length scales, too. We shall not go into further details
and refer the reader to the `polarisable vacuum representation' of GR by \cite{Put:99},
see also \cite{Unz:05}. It is however at least an open question if GR can be formulated
by a scalar VSL theory, instead of a 10-component metric! I shall not enter the fruitless
question whether this is `simple' or not - it's up to you whether you consider this an
approach worth thinking about or share the above criticism: what a pity
that Einstein in 1911 could not make use of check-lists like \cite{Ell:07} -
maybe he had stopped to develop weird theories about a curved spacetime...

\paragraph{Lorentz invariance.} Such useful methodic guidelines would also have
prevented Einstein from being in conflict with special relativity and remaining so blatantly
unfamiliar with underlying principles of his own work - I wonder if this is the message the
reader should learn from \cite{Ell:07}, sec.~4.
In 1911, Einstein wrote:
\bq
The constancy of the velocity of light can be maintained only insofar
as one restricts ... to ... regions with constant gravitational potential...
\eq
Indeed, such a VSL theory would require considerable reformulation on a technical
level, but there is little doubt that this can be done as a matter of principle, as
long as the local $c$ is the limiting velocity. In continuum
mechanics it is well-known that special relativistic effects show up \cite{Unz:00},
 and a variation of the (corresponding) speed of transversal sound
arises naturally. Further clarifying explanations are given by Dicke \cite{Dic:57}
and Ranada \cite{Ran:04a}.

\subsection{The Conditions a good physical theory has to satisfy }
The necessity of compatibility with special relativity or
Lorentz-invariance of a theory stressed by \cite{Ell:07} is one of
the most basic requirements new proposals have to satisfy in order
to be taken seriously. Other requirements would be the possibility
of a Lagrangian formulation, satisfying the equivalence principle,
the agreement with general relativity, with quantum
electrodynamics, to be renormalizable etc. All these are nice
properties of successful physical theories.  Setting up guidelines
for the development of possible new ideas however does not provide
any real progress. At the very end, there is only one really
significant test: theories have to be in agreement with experiment
and the observations.  This basic need sometimes is forgotten,
maybe because nowadays scientific methodology is quite dominated by
mathematical constructions like string theory which have difficulties
to get the link to experiments \cite{Smo:06, Woi:06}.

\paragraph{Parameters, fields, and simplicity}

Not really a requirement, but a hint to a good physical theory is
simplicity and economy of concepts; this is sometimes called
`Occams razor'. Economy is hardly any more a property of physics'
standard models; particle physics has about 20 freely adjustable
parameters, and cosmology, constrained to digest new data, is
currently producing new ones\cite{Unz:07}. In physics we have a
dilaton field, an inflaton field, a Higgs field, dark matter, the
cosmological constant became a quintessence field, not to mention
numerous proposals with a shorter life time. Given that all that
is undoubtedly convincing, is then anything else but $c \equiv 1$
too complicated physics~?

\paragraph{The need and the fear to change equations.}

Paradoxically, postulating exotic new fields does usually little
harm to the standard models, while the speed of light has a
dominant role in various fields of theoretical physics
\cite{Eli:03}. VSL has obviously to
consider the influence on other fields since changing $c$ in one
context only would be a rather weird and fruitless trial. On the
other hand, despite all technical difficulties that may arise,
the ultimate 
test remains the agreement with experiment, and due to the usual
minute deviations a VSL causes we cannot expect that the corresponding
observations become visible in all facets
simultaneously. Of course, as \cite{Ell:07} states, if one changes one equation, one
has to change many ones, but this elucidates also the psychological problem
that may arise: for somebody who has written a book full of
formulas containing $c$, any VSL proposal becomes a nightmare.

\paragraph{Physicists in the ptolemaic period}
may have felt similarly when hearing about the earth being in motion.
For somebody living in the 17th century,  surely it wasn't easy to
get familiar with such a counterintuitive fact. But asking `if $c$ is variable why
don't we measure a change~? is like asking
`If earth is moving around the sun, why isn't there a strong wind blowing due to that
motion~?'. Galilei responded:
\bq
`Close yourself with a friend in a possibly large room  below deck
in a big ship. [...] ...of all appearances you will not be able to deduce
a minute deviation...' \cite{Sin}
\eq
It is a fact that the dynamics of a system  we are part of  are sometimes
quite difficult to detect; for conservative people it is then
much more evident to stick to the stationary image 
that maintains stability, and it is easy to ask questions like
`How do you express formula xyz of the common formalism in the new
formalism~?' \fo{\cite{Ell:07}, implication~2.} Try to formulate
the deferrents and epicycles of the Ptolemaic world view in
Newton's language - what a medieval torture!
Beholding 
the present situation of physics from a historical perspective (e.g. with the
excellent popular book \cite{Sin}) may be helpful to get a distance to the
belief in the validity and generality of our present theories we use
for starry-eyed extrapolations.

\subsection{Outlook}
To believe or not to believe if VSL is a promising approach in physics is not a scientific
question; if one does not, he is free to continue the work he finds interesting to do.
Thus it is not necessary to develop toolkits enabling a critique 
of any VSL paper \cite{Ell:07}. We certainly do not need proofs
that VSL cannot be an adequate approach,
because (1) such a proof does not exist and (2) science has never advanced with
such proofs. Neither we do need warnings that anything else that the
standard model is dangerous, and statements like `if you think about
anything else than the standard model, you have to
deliver a complete solution immediately' (\cite{Ell:07}, implication~5).

Physics needs the close link to experiments and observations and a freedom of ideas and methods.

\end{document}